# An auto-initializing algorithm to determine electronic ground states from random Slater determinants

Hao Zhang [1] ✉ & Matthew Otten [1,2] ✉

Accurate quantum many-body calculations often depend on reliable reference states or good human-designed ansätze, yet such knowledge can become unreliable in hard problems like strongly correlated systems. Here we show that high-accuracy many-body ground states can be discovered directly from random Slater determinants. We introduce the Trimmed Configuration Interaction (TrimCI) method, which self-refines by iteratively expanding and trimming configurations in the variational space. TrimCI matches state-of-the-art accuracy using orders of magnitude fewer determinants and computer time: $10^6$-fold fewer than a recent quantum computing calculation on an iron-sulfur cluster, $10^5$-fold fewer than a leading classical method on a larger iron-sulfur cluster in nitrogenase, and over 99% of the ground-state energy of a strongly correlated lattice system from $10^{-28}$ of its Hilbert space, exceeding quantum Monte Carlo accuracy by orders of magnitude in some regimes. TrimCI thus provides an auto-initializing, accurate and highly efficient framework for quantum many-body systems, whose compact explicit wavefunctions enable direct and rapid evaluation of observables.

A central theme in quantum many-body physics and quantum chemistry is to compute accurate ground states with tractable resources. Over the past decades, transformative advances have steadily pushed this frontier[1–4]. Tensor-network methods[5–15], exemplified by DMRG[5–7] and extensions to higher dimensions[8,11], exploit entanglement structure to compress otherwise exponentially large Hilbert spaces. In the Quantum Monte Carlo (QMC) family[16–25], such as auxiliary-field variants and variational variants, stochastic and sampling techniques are used to get an accurate estimate of ground states. In determinant space, selected configuration interaction (selected-CI)[26–35] selects determinants with strong couplings to a good reference state like the Hartree-Fock state. More recent approaches, e.g., variational quantum eigensolvers (VQE)[36–42] and neural quantum states (NQS)[43–47] pursue expressive variational ansätze while leveraging modern hardware (GPUs)[48,49] and emerging quantum processors (QPUs)[50–60]. Together these methods delineate today's state of the art.

Despite the challenge of exponentially large Hilbert spaces, a unifying strategy behind many successes of these methods is to use explicit knowledge or assumptions about the system to restrict the search[15,61,62]: predesigned variational ansätze, guiding or trial wavefunctions, or favorable initializations. For example, matrix-product states (MPS)[6], projected-entangled-pair states (PEPS)[8,15] and fermionic tensor product state (fTPS)[11,14] use ansätze with limited entanglement[12] to approximate the ground states. Auxiliary-field QMC (AFQMC)[21] uses a good trial wavefunction[20,63] to guide importance sampling. Variational QMC (VMC)[22] and related methods benefit from good initializations that alleviate autocorrelation in samples. Methods like Selected-CI and many VQE workflows often start from Hartree-Fock[30,36,38]. These designs work very well when reliable explicit knowledge is available. However, in strongly correlated regimes, with high entanglement, competing orders, or poorly understood ground states, such explicit knowledge is often unavailable or difficult to obtain[2,64,65]. Consequently, tensor-network methods must employ larger bond dimensions[66], variational quantum circuits require deeper and more complex architectures[67], and neural quantum states demand additional layers to maintain accuracy. Sampling-based approaches, in turn, often need far more samples[68] to recover contributions missed by the trial state or to escape incorrect energy basins[69], delaying the discovery of configurations that dominate the exact ground state. Yet, a practical lesson from these existing methods emerges: if one can identify the important part of Hilbert space that contributes most to the ground state, then accurate solutions may still be accessible.

[1]Department of Physics, University of Wisconsin—Madison, Madison, WI, USA. [2]Department of Chemistry, University of Wisconsin—Madison, Madison, WI, USA. ✉e-mail: hao.zhang.quantum@gmail.com; mjotten@wisc.edu

Motivated by this insight, we explore an auto-initializing approach, termed Trimmed Configuration Interaction (TrimCI). It requires no explicitly provided problem-specific knowledge, no predesigned variational subspace, no guiding or trial wavefunction, and no chosen starting point. Surprisingly, it can efficiently find the important part of the Hilbert space by itself and produce state-of-the-art accuracies. For hard problems, it can even find a much better core set than a human-designed one.

TrimCI operates on the graph whose nodes are many-body Fock basis states (Slater determinants for fermions) and whose edges correspond to nonzero Hamiltonian couplings $H_{ij}$. Starting from a provisional core set consisting of random basis states, it alternates between two operations (see Fig. 1(a, b)): expansion, which augments the core by adding neighbors connected by sufficiently large couplings $|H_{ij}|$; and trimming, which removes negligible basis states by first diagonalizing randomized blocks of the core and then performing a global diagonalizing step on the surviving set. Iterating expansion and trimming drives the core set to cover the most important regions of the graph that harbor the dominant basis states. At each iteration, TrimCI solves the many-body problem within the core set, thereby providing a variational upper bound that monotonically improves as the core is refined.

Extensive tests across challenging molecular and lattice benchmarks support the effectiveness of TrimCI. For strongly correlated molecules, e.g., Chromium dimer[70] (36 spatial orbitals and 48 electrons), [4Fe-4S] cluster[71] (36o, 54e) and P-cluster in nitrogenase[62] (73o, 114e), TrimCI matches selected-CI accuracy while using notably $10^2$-fold to $10^5$-fold fewer determinants, see Fig. 1c. On the Hubbard model[2,15,21,64,65], TrimCI efficiently identifies the correlated ground-state basin (Fig. 1d), recovering about 99%-99.99% of the ground-state energy with very compact determinant sets. On the $4 \times 4$ lattice at half-filling with $U = 2$, our results surpass the accuracy of AFQMC benchmarks reported in Ref. 21. On the largest lattice, the $8 \times 8$ Hubbard model at half-filling with $U = 2$, TrimCI attains 99% of the ground-state energy using only $3 \times 10^8$ determinants out of a Hilbert space of order $10^{36}$. Across $4 \times 4$, $6 \times 6$, and $8 \times 8$ lattices, the fraction of Hilbert space actually used by TrimCI decreases with system size, suggesting a vanishing fraction in the thermodynamic limit.

In this work, we show that TrimCI constructs a compact and explicit ground-state wavefunction, providing a highly efficient route for accurate many-body computation and enabling immediate evaluations of observables with resources.

## Results

### Motivation

The Hilbert space of a quantum many-body system can be viewed as a weighted graph $G = (V, E, w)$, where each node $|i\rangle \in V$ represents a Fock basis state, and each edge $(i, j) \in E$ connects two basis states coupled by a nonzero Hamiltonian element $H_{ij}$. The edge weight $w_{ij} = |H_{ij}|$ characterizes the coupling strength between the two nodes. In this picture, optimizing the wavefunction is equivalent to performing a diffusion process over this graph, guided by the connectivity and heterogeneous strengths of $H_{ij}$.

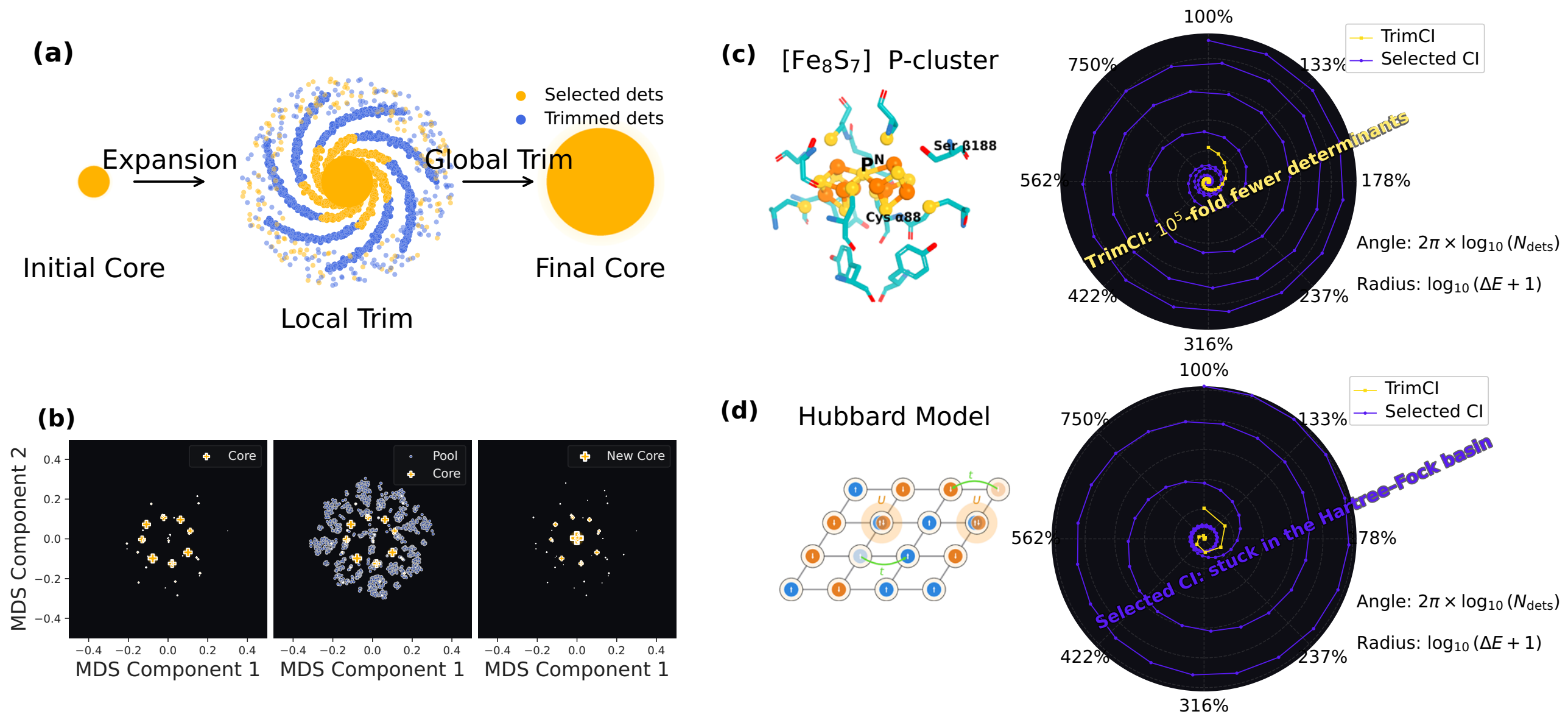

**Fig. 1 | Overview and efficiency of the Trimmed Configuration Interaction (TrimCI) algorithm. a** Schematic workflow. Starting from a random initial core of determinants, TrimCI alternates between expansion, which adds connected determinants via Hamiltonian couplings, and trimming, which removes unimportant ones, ultimately converging to a compact final core containing the dominant configurations of the ground state. One schematic iteration is illustrated. **b** Example of a TrimCI iteration using real data. Multidimensional scaling (MDS) visualizations project the high-dimensional determinant space onto two dimensions, with marker size indicating the magnitude of the wavefunction coefficients. The plots show the evolution from an initial core to an expanded pool and finally to a refined new core, illustrating how TrimCI autonomously identifies the most important determinants within the Hilbert space. **c** Application to the $[Fe_8S_7]$ P-cluster; the structure shown was rendered with PyMOL from Protein Data Bank entry 3MIN[99]. TrimCI achieves the same accuracy as selected-CI while using over $10^5$-fold fewer determinants, as visualized by the polar efficiency spiral, where the radius encodes the energy deviation $\log_{10}(\Delta E + 1)$ relative to a reference point and the angle represents the logarithmic determinant cost $2\pi \log_{10}(N_{\text{dets}})$. **d** Application to the two-dimensional Hubbard model. TrimCI efficiently discovers the correlated ground-state basin, whereas selected-CI remains trapped near the Hartree--Fock minimum, highlighting TrimCI's ability to escape local basins and identify globally optimal determinant subsets.

The goal of TrimCI is to identify a compact core set $C \subset V$ that captures the dominant weight of the ground-state wavefunction,

$$|\Psi_0\rangle = \sum_{i \in V} c_i |i\rangle, \quad (1)$$

such that diagonalization of the projected Hamiltonian

$$H_C = P_C H P_C,\; P_C = \sum_{i \in C} |i\rangle\langle i|, \quad (2)$$

yields a variational energy

$$E_C = \min_{\psi \in \mathrm{span}(C)} \frac{\langle \psi | H_C | \psi \rangle}{\langle \psi | \psi \rangle}, \quad (3)$$

which satisfies $E_C \geq E_0$ and converges to the exact ground-state energy $E_0$ as $C$ is refined. Efficient iterative solvers such as Davidson[72] can be used for this diagonalization. In what follows, we focus on fermionic systems, where the nodes correspond to Slater determinants.

At first glance, due to the exponentially large size of the Hilbert space, identifying a compact core set without a reference state (e.g., Hartree-Fock) appears hopeless. However, the underlying graph possesses rich structural features that encode the physics of the system, which can be exploited algorithmically. For a fermionic Hamiltonian

$$H = \sum_{pq} t_{pq} c_p^\dagger c_q + \frac{1}{2} \sum_{pqrs} v_{pqrs} c_p^\dagger c_q^\dagger c_s c_r, \quad (4)$$

two determinants differing by at most two orbital occupations are directly connected through $H_{ij}$[73]. Here $c_p^\dagger$ and $c_q$ are fermionic creation and annihilation operators, $t_{pq}$ are one-electron integrals, and $v_{pqrs}$ are two-electron integrals. Consequently, any two determinants (within the same connected component) are separated by a short path containing at most $O(m)$ hops, where $m$ is the number of orbitals (single particle basis states). Moreover, realistic Hamiltonians, e.g., in molecular and lattice systems, exhibit strongly heterogeneous couplings, spanning several orders of magnitude. These heterogeneities form major routes that efficiently bridge random determinants to those with large ground-state amplitudes. TrimCI exploits such routes to uncover the important core subgraph supporting the ground state.

## Algorithm

The algorithm starts with a relatively small random initial set $C_0$ of determinants that satisfy the required particle number and spin constraints (correct $S_z$ value). Two complementary stages alternate to refine the core.

(1) Expansion step: the core set $C$ expands into a pool set $P$ by including determinants connected to the core through sufficiently large couplings $H_{ij}$. A simple but computationally efficient heuristic rule may be used:

$$P = C \cup \{ i \,|\, |H_{ij} c_j| > \theta, j \in C \}, \quad (5)$$

where $c_j$ is the coefficient of det $j$ from the previous iteration. Note there is no summation over $j$ here for fast calculation[30]. $\theta$ is a dynamically adjusted threshold ensuring a reasonably large pool (typically we choose $|P| = 10|C|$ or $|P| = 100|C|$). This step enlarges the search space along physically significant edges of the graph.

(2) Trimming step: the trimming stage eliminates unimportant determinants from the expanded pool $P$ in two successive substeps. (2a) Local trimming: the pool $P$ is first divided into several, random groups $R_\alpha$. Within each group, diagonalization of projected Hamiltonian $H_{R_\alpha}$ produces their approximate local coefficients, and only top $k_a$ determinants are retained for the next stage. This randomized trimming quickly discards evidently negligible configurations while preserving representatives from diverse regions of the graph. (2b) Global trimming: the surviving determinants from all groups are then collected into a reduced pool $P'$, on which a single global diagonalization of $H_{P'}$ is performed to obtain more accurate coefficients $c_i^{\mathrm{next}}$. The next core $C^{\mathrm{next}}$ is chosen by selecting the top-$k_b$ determinants with the largest amplitudes $|c_i^{\mathrm{next}}|$,

$$C^{\mathrm{next}} = \{ i \,|\, |c_i| \in \text{top} - k_b,\; i \in P' \}. \quad (6)$$

This controlled two-level trimming effectively refines the core set, leading to a nearly monotonic improvement of the variational energy towards the ground-state limit. See Suppl. Note 3 of the Supplementary Information for a detailed study of the trim mechanism.

The parameters $k_a$ and $k_b$ can be dynamically tuned to improve efficiency. In the first iteration, because of random initialization, a smaller ratio of determinants is kept for the next core. We observed that after the early iterations, TrimCI can quickly find a tiny core set consisting of the most important determinants in the ground state. Thus in the late iterations, one may skip the trimming step and perform only expansions and global diagonalization for coefficient updates. In this sense, TrimCI can be regarded as a more general algorithm than selected-CI methods.

To improve efficiency, TrimCI can be executed with several random initializations (in parallel) for the first few iterations and then the best-performing run is chosen for longer iterations. This ensemble strategy accelerates convergence and enhances robustness, increasing the likelihood of discovering the globally optimal core. In practice, we found that 64 independent initializations running up to 100 determinants is a good default. See Suppl. Notes 1 and 2 for details.

After finding a good core set, TrimCI can be incorporated into other existing workflows. Orbital optimization[33] can be performed to further increase the compactness of the wavefunction. We found that with the core set produced by TrimCI, orbital optimization becomes much more effective. Perturbative corrections and extrapolations[32] can also be applied to obtain a better energy estimate with only small additional computational resources. A complete TrimCI workflow example is given in Suppl. Table 1.

## Molecular systems

We first benchmarked the performance of TrimCI on several challenging molecular systems[62,70,71], as summarized in Fig. 2 and Table 1. These systems range from the strongly correlated chromium dimer to the highly entangled iron-sulfur clusters. $Cr_2$ uses Hartree–Fock canonical orbitals; the iron-sulfur clusters ([2Fe-2S], [4Fe-4S], and the nitrogenase P-cluster) use localized orbitals from the literature[62,71]. Among them, the P-cluster in nitrogenase[62] is of particular importance: it serves as a key catalytic component of biological nitrogen fixation, and due to its large active space (73*o*, 114*e*) and complex multi-center exchange couplings, it remains one of the most demanding and scientifically significant frontiers for accurate quantum many-body computation. In the following, we use the same orbitals to compare with selected CI, to isolate the effect of the orbital basis. See also Suppl. Note 1 for comparisons with scrambled orbitals.

Panels (a), (d), and (g) of Fig. 2 compare the convergence of variational energy with the number of determinants for TrimCI and the state-of-the-art Selected CI method (SHCI[30–33]). Across all systems, TrimCI achieves the same accuracy using $10^2$-fold to $10^5$-fold fewer determinants.

For $Cr_2$ (36o, 48e, STO-3G basis, $d$ = 2.5 Å), TrimCI reaches the energy of −2064.5 Ha with only $1.0 \times 10^3$ determinants, while SHCI requires $6.8 \times 10^4$. In the [4Fe-4S] cluster (36o, 54e), the efficiency gain increases to $7.5 \times 10^2$, and for the nitrogenase P-cluster (73o, 114e), it reaches over $10^5$: 100 determinants of TrimCI to reach the energy that SHCI takes $10^7$–$10^8$ determinants. Even for smaller systems such as [2Fe-2S] with (20o, 30e), TrimCI remains roughly 40 × more efficient. As summarized in Table 1, the performance advantage of TrimCI becomes more pronounced with increasing correlation strength, highlighting its effectiveness and robustness in the multireference and strongly correlated regime. Comprehensive benchmarking including PT2 corrections and state-of-the-art calculations of $Cr_2$ in the cc-pVDZ-DK basis with CAS(28e, 76o) are reported in Suppl. Note 6, with wavefunction-level analysis of $Cr_2$ in Suppl. Note 4.

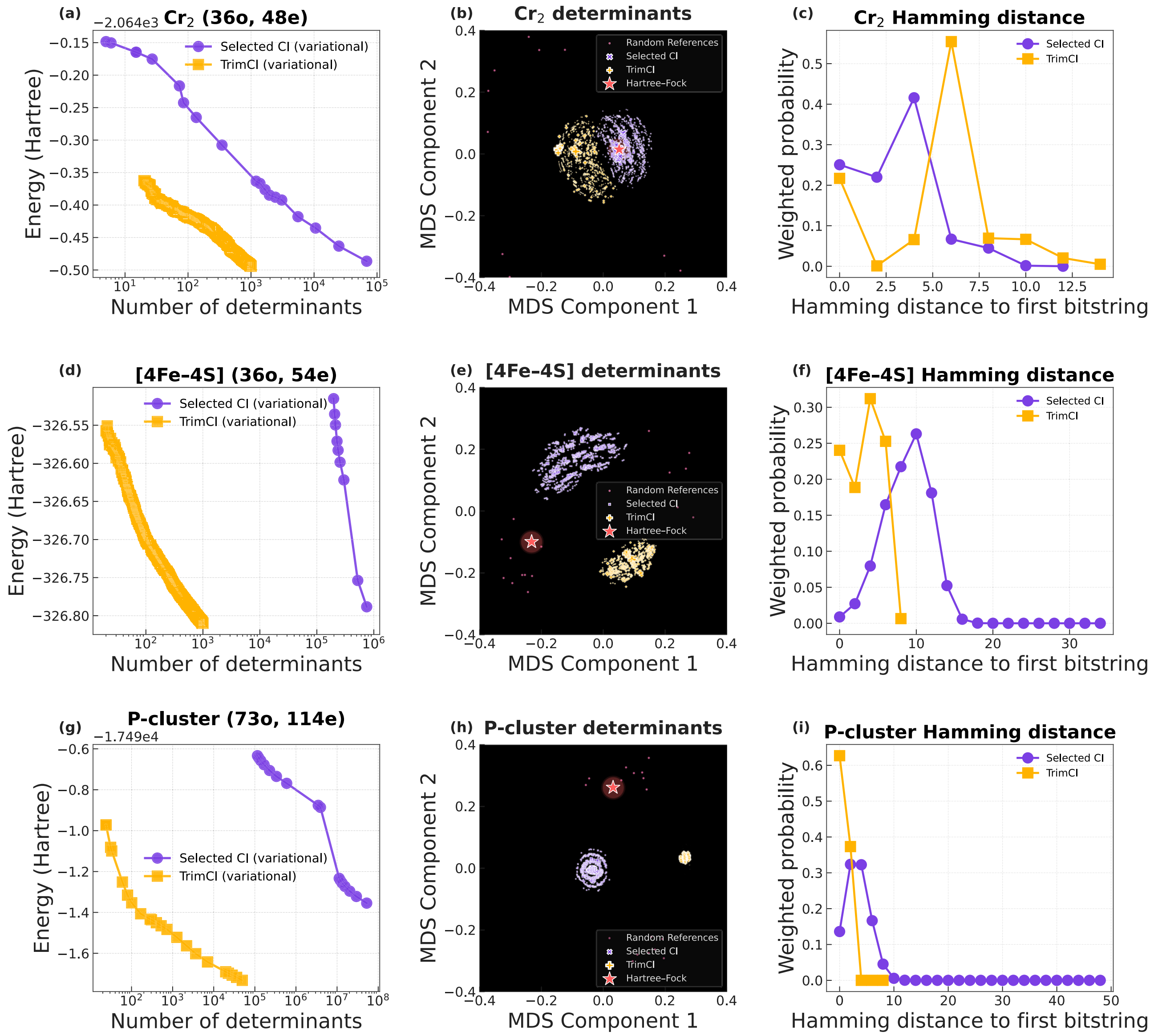

**Fig. 2 | TrimCI vs. selected-CI on molecular benchmarks. a**, **d**, **g** Energy convergence for $Cr_2$ at $r = 2.5$ Å (36o, 48e), [4Fe-4S] (36o, 54e), and P-cluster (73o, 114e). TrimCI reaches comparable accuracy with $10^2$–$10^5$ × fewer determinants. **b**, **e**, **h** MDS embeddings of the largest variational wavefunctions in (a,d,g). Each point is a Slater determinant with marker size ∝ coefficient magnitude. TrimCI (gold) forms compact manifolds near the physically relevant region; Selected-CI (violet) spreads around HF (red star) or wrong basins. **c**, **f**, **i** Hamming-distance distributions (i.e., 2× excitation level) from the leading determinant. TrimCI exhibits narrower distributions, indicating more efficient identification of the entanglement core. Source data are provided as a Source Data file.

To gain intuition into this distinct efficiency of TrimCI, we projected the many-body wavefunction into a two-dimensional plane using multidimensional scaling (MDS)[53], as shown in panels (b), (e), and (h) of Fig. 2. Each point corresponds to a determinant, with the size of the marker proportional to its coefficient magnitude. The visualizations reveal that SHCI and TrimCI explore fundamentally different regions of Hilbert space. For $Cr_2$, SHCI remains concentrated near the Hartree-Fock configuration, while TrimCI immediately identifies a distinct region away from it, the region containing the important weight of the exact ground state. For the [4Fe-4S] and P-clusters, both methods eventually move beyond the Hartree-Fock region, but SHCI expands to an unfavorable region due to a poor initial reference, whereas TrimCI directly focuses on the energetically more relevant basin due to its iterations of expansion and trimming. This difference in exploration dynamics explains the $10^2$-$10^5$ × reduction in determinant count required to reach the same energy accuracy.

Panels (c), (f), and (i) of Fig. 2 show the weighted probability distributions of determinants as a function of Hamming distance from the det $i_{max}$ with the largest coefficient magnitude (i.e., the number of differing orbital occupations, equal to 2 × the excitation level),

$$P(d) = \sum_{i:\,\mathrm{Ham}(i, i_{max}) = d} |c_i|^2. \tag{7}$$

TrimCI wavefunctions exhibit a narrower distribution, with effective widths roughly two times smaller than SHCI. This compactness reflects

**Table 1 | TrimCI benchmark results on representative strongly correlated molecular systems**

| System | (Orbitals, Electrons) | TrimCI Energy (Ha, $10^3$ dets) | #Dets of SHCI | Efficiency |
|---|---|---|---|---|
| P-cluster in nitrogenase | (73o, 114e) | −17,491.50 | >5.3 × $10^7$ | >53,000× |
| [4Fe-4S] cluster | (36o, 54e) | −326.810 | 7.5 × $10^5$ | 750× |
| $Cr_2$ ($d$ = 2.5 Å, STO-3G) | (36o, 48e) | −2064.494 | 6.8 × $10^4$ | 68× |
| [2Fe-2S] cluster | (20o, 30e) | −116.459 | 3.9 × $10^4$ | 39× |
| $Cr_2$ ($d$ = 3.0 Å, STO-3G) | (36o, 48e) | −2064.565 | 8.5 × $10^3$ | 8.5× |
| $Cr_2$ ($d$ = 2.0 Å, STO-3G) | (36o, 48e) | −2064.503 | 6.2 × $10^3$ | 6.2× |

Each entry shows the system, TrimCI energy in Hartree (Ha) obtained with $10^3$ determinants, the number of determinants used by SHCI to reach comparable accuracy, and the resulting efficiency ratio.

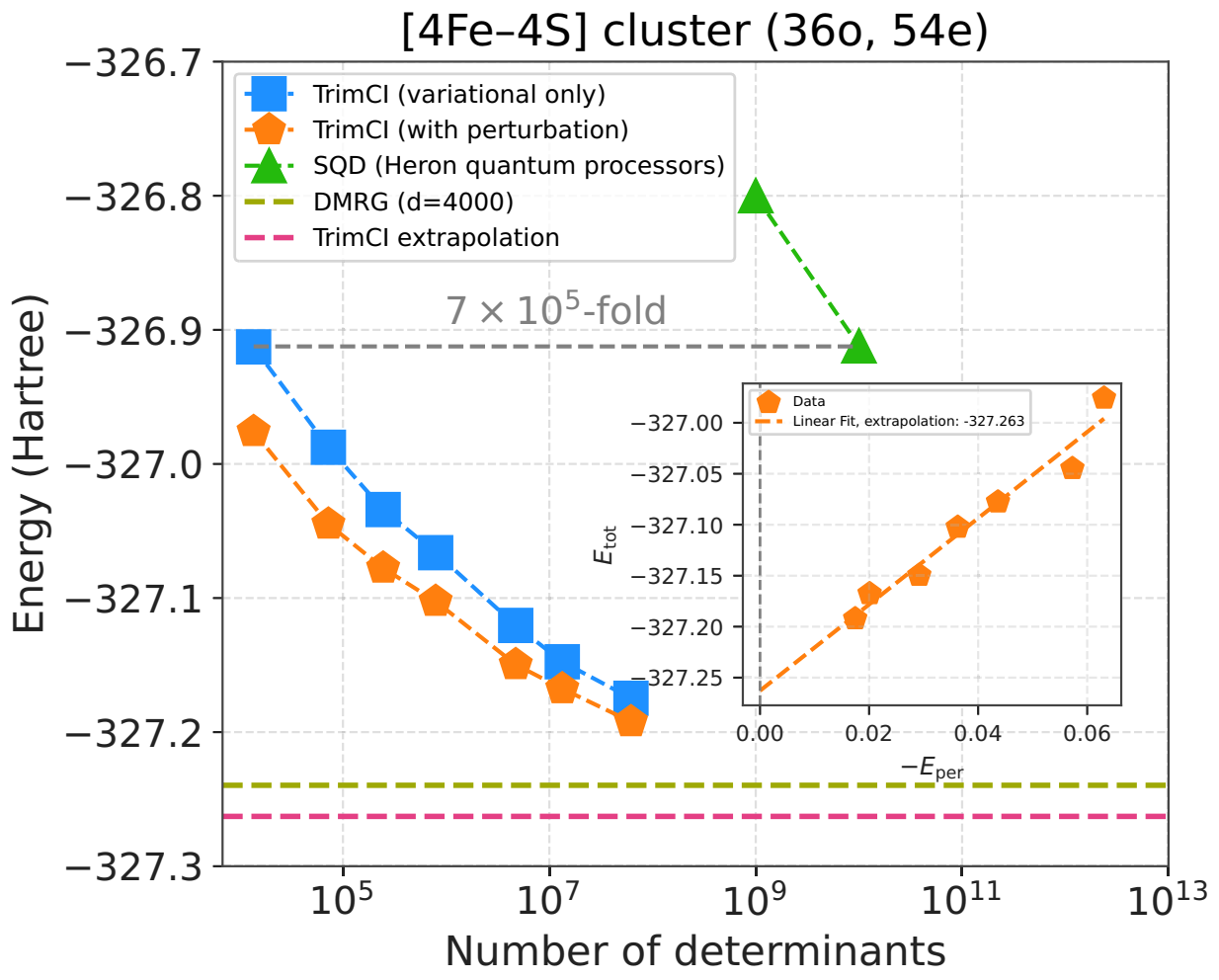

**Fig. 3 | Ground state energy of [4Fe-4S] cluster with TrimCI compared to SQD.** TrimCI achieves the same energy ($E = -326.9127$ Ha) as the SQD method, which used $10^{10}$ determinants and $10^7$ CPU-h, with only $1.3 \times 10^4$ determinants and 5 CPU-h. This corresponds to over $7 \times 10^5$-fold efficiency in reaching the same accuracy. The TrimCI variational and perturbatively corrected energies show smooth convergence. Source data are provided as a Source Data file.

that TrimCI isolates a physically meaningful core subspace where the dominant configurations reside, leading to faster convergence and greater numerical efficiency.

Taken together, these results demonstrate that TrimCI can indeed identify a much more important and compact region of the Hilbert space, thereby improving its efficiency by several orders of magnitude.

## Ground state of the [4Fe-4S] cluster

To further benchmark TrimCI with existing methods, we compute the ground state energy of the [4Fe-4S] cluster, a complicated iron-sulfur system whose strongly correlated nature presents a significant challenge for quantum many-body algorithms. A recent study[74] applied Sample-based Quantum Diagonalization (SQD), using $10^{10}$ Slater determinants generated by a parameterized quantum circuit on the 156-qubit Heron processor, and achieved a variational energy of $E = -326.912$ Ha. The associated classical processing required 152,064 nodes and 7.3 million CPU cores, running for at least 15 min per iteration over 10 iterations, exceeding $10^7$ CPU-h in total.

As shown in Fig. 3, TrimCI surpasses this result while using orders of magnitude fewer determinants. It reaches −326.810 Ha using only $10^3$ determinants and matches the SQD energy of −326.9127 Ha with merely $1.3 \times 10^4$ determinants. This is six orders of magnitude fewer than SQD. Furthermore, with TrimCI's orbital optimization enabled, the full workflow reaches $N_{\rm det} = 5.12 \times 10^9$ determinants and an extrapolated energy of −327.2441 Ha, within 0.2 mHa of the large-scale UDMRG extrapolation on the same broken-symmetry state (see Suppl. Notes 7 and 8). We also refer the reader to refs. 75,76 for more accurate and detailed results on this system.

Crucially, the variational TrimCI calculation requires vastly fewer computational resources. For example, the −326.9127 Ha result was obtained on a local laptop using 20 CPU cores in 15 min, amounting to only 5 CPU-h, while SQD consumed over $10^7$ CPU-h in addition to quantum resources. TrimCI thus achieves $10^6$-fold efficiency in both memory and compute.

The main reason behind this large efficiency gain is that TrimCI does not rely on a good starting wavefunction. Methods such as SQD encode problem-specific heuristics into parameterized quantum circuits and depend on good initial states such as Hartree-Fock. While such priors may be very successful for some systems, they can become inaccurate or even misleading in strongly correlated settings, resulting in vast algorithmic overhead. TrimCI takes a completely different approach: it begins from randomized determinants and navigates the Hilbert space solely through the Hamiltonian graph structure, without imposing any bias of the ground state.

Through its expansion-trimming cycles, TrimCI autonomously discovers a highly compact and physically expressive core subspace. Expansion explores new determinants connected by the Hamiltonian, while trimming removes unimportant configurations and rapidly concentrates amplitude onto the essential many-body structures. This automated, trimming-driven process is what enables TrimCI to reach accurate quantum many-body energies with such efficiency, achieving results that required millions of CPU hours and quantum hardware by SQD, now on a laptop within hours. We further show in Supplementary Note 7 that, even starting from a fully scrambled orbital basis on the same [4Fe-4S] Hamiltonian, the full TrimCI workflow recovers an orbital basis on which the variational expansion converges rapidly and extrapolates to the FCI limit.

## Lattice systems

Having established the performance of TrimCI on molecular systems, we next turn to lattice Hamiltonians, where the challenge arises not from chemical complexity but from strong correlation and collective quantum fluctuations. In particular, we focus on the two-dimensional Hubbard model[2,15,21,64,65,77] as a paradigmatic platform for benchmarking many-body algorithms. All Hubbard calculations use Hartree–Fock canonical orbitals. The results are summarized in Fig. 4 and Table 2.

For the 4 × 4 Hubbard model ($U$ = 2, half-filling) with periodic boundary conditions, TrimCI results are compared against exact diagonalization (full configuration interaction (FCI)) energy[78]. As shown in Fig. 4a, the relative energy error of TrimCI variational energy exhibits a clear power-law decay with the number of determinants, indicating systematic convergence toward the exact limit. Only $4.2 \times 10^2$ determinants are needed to reach within 1% error rate, and $8.7 \times 10^5$ determinants already achieve a relative error of $10^{-4}$, corresponding to merely 0.5% of the total Hilbert-space size ($1.7 \times 10^8$).

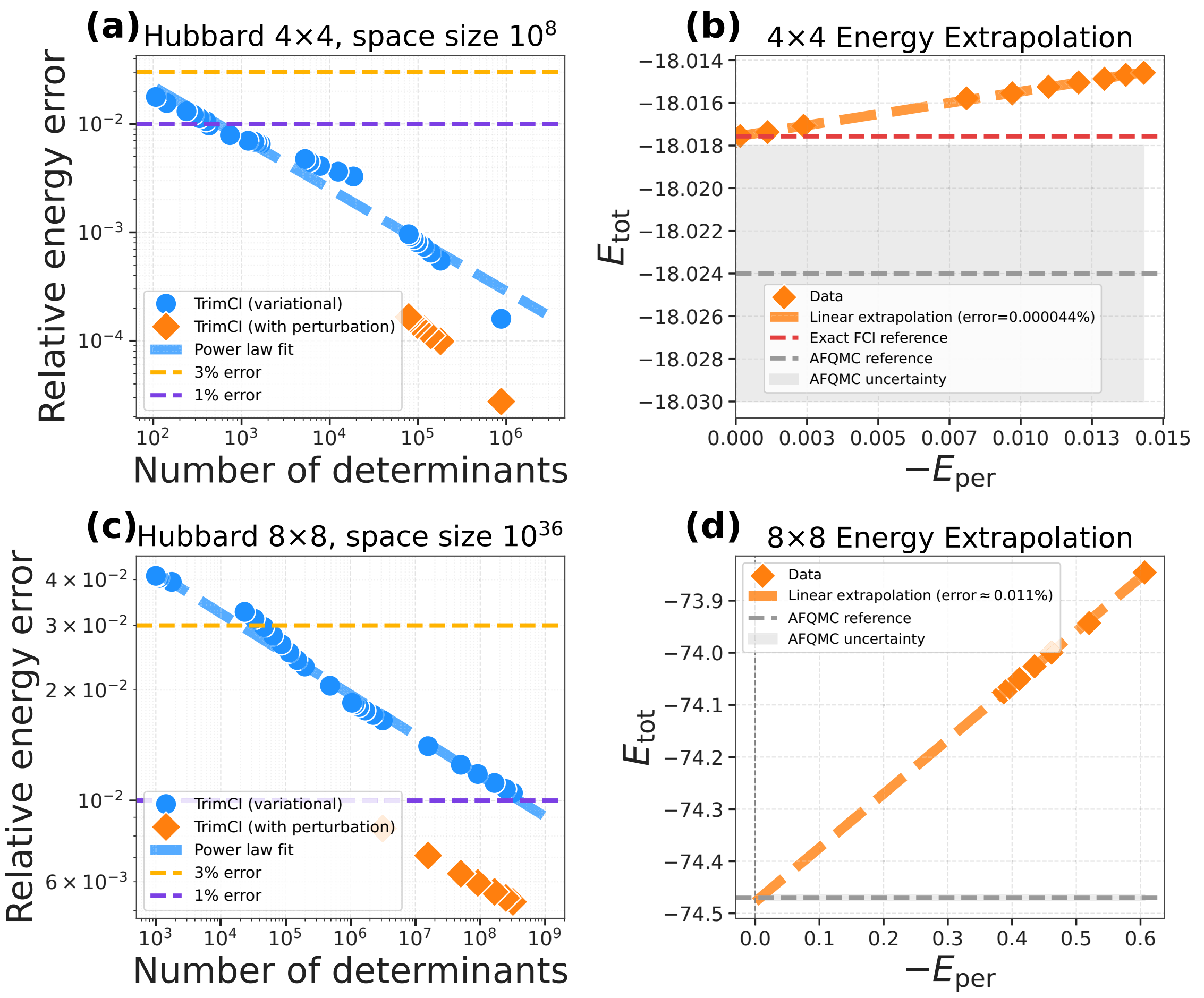

**Fig. 4 | TrimCI performance on the two-dimensional Hubbard model.** All energies are given in units of the hopping amplitude $t$, and the interaction strength is quoted as $U/t$. **a** Relative energy error versus number of determinants for the 4 × 4 lattice with periodic boundary condition ($U = 2$, half filling), compared to exact FCI results. Dashed blue lines show power-law fits; orange markers include perturbative corrections. **b** Linear extrapolation of total energy $E_{tot} = E_{var} + E_{per}$ against $-E_{per}$, converging to the exact FCI reference within a relative error of $4 \times 10^{-7}$; AFQMC reference and uncertainty are shown in gray. **c** Scaling for the 8 × 8 lattice with periodic boundary condition ($U = 2$, half-filling); TrimCI achieves 97-99% accuracy with $10^4$–$10^8$ determinants in a space of $10^{36}$ determinants. **d** Energy extrapolation for the 8 × 8 lattice, where the extrapolated energy deviates from the AFQMC reference by only 0.011%. Source data are provided as a Source Data file.

We further include the perturbative correction (orange markers), obtained by incorporating determinants connected to the TrimCI core. This correction systematically improves the energy estimation with only modest computational overhead. To quantify the extrapolation accuracy, we plot $E_{tot} = E_{var} + E_{per}$ versus $-E_{per}$ in Fig. 4(b). A linear extrapolation yields an energy that matches the exact FCI value almost perfectly, with a relative error of just $4 \times 10^{-7}$. This result is notably more accurate than the AFQMC reference energy from Qin et al.[21] (gray dashed line) which has a relative error of $4 \times 10^{-4}$.

This discrepancy highlights the different convergence properties of the methods. While AFQMC is statistically exact in principle, its practical accuracy is limited by the quality of the trial wavefunction and the statistical error, which scales slowly as $1/\sqrt{N_{samples}}$. To match the $10^3$-fold higher accuracy of the TrimCI extrapolation, AFQMC would require roughly $10^6$ times more samples. TrimCI, in contrast, takes a different approach, achieving its accuracy by systematically compressing the Hilbert space into a core set that captures the dominant ground-state contributions.

For the 8 × 8 Hubbard model ($U = 2$, half-filling) with periodic boundary conditions, the Hilbert-space dimension grows to $3.4 \times 10^{36}$. Here, the AFQMC energy from Qin et al.[21] serves as the reference. As shown in Fig. 4c, TrimCI maintains power-law convergence across five orders of magnitude in determinant count, suggesting it is converging correctly. With $4.4 \times 10^4$ determinants, the energy reaches 97.0% of the AFQMC reference; with $2.5 \times 10^8$ determinants, the accuracy improves to 98.9%. Incorporating the perturbative correction further reduces the energy error to 0.54%.

Extrapolation, shown in Fig. 4d, demonstrates that the total energy $E_{tot}$ linearly approaches the AFQMC reference, with a final deviation of only 0.011%. This confirms that TrimCI captures the core of the ground state efficiently even in a space of $10^{36}$ configurations with a core of only $10^8$ determinants. Based on the power law trend in panel (c), we can continue to increase the accuracy by growing the core. See Suppl. Note 5 for order parameter analysis.

Other results are summarized in Table 2. We compare the results with different system sizes with same conditions ($U = 2$, half-filling, periodic boundary condition). For lattice of size $N = 4 \times 4, 6 \times 6, 8 \times 8$, the fraction $r$ of the Hilbert space required to reach 99% accuracy of the variational energy is found to be $2.5 \times 10^{-6}$, $3.6 \times 10^{-16}$, and $1.5 \times 10^{-28}$, respectively. The data support a linear scaling with logarithmic form,

$$\log_{10} r = -0.462N + 1.559, \tag{8}$$

with $R^2 = 0.9991$, indicating an excellent linear correlation. Since $r = N_{dets}/N_{total}$, where $N_{dets}$ is the number of determinants in the TrimCI core and $N_{total}$ is the total Hilbert space dimension, and $\log_{10} N_{total} \approx 2\log_{10} 2 \cdot N$ for half-filled Hubbard (to leading order in $N$), the actual

**Table 2 | TrimCI benchmark results on Hubbard lattices**

| System | | | | | TrimCI Accuracy Levels | | | | Extrapolation | | |
|---|---|---|---|---|---|---|---|---|---|---|---|
| Lattice | $U/t$ | $N_{total}$ | FCI Ref. ($t$) | AFQMC Ref. ($t$) | Ratio | Energy ($t$) | $n_{dets}$ | Trim | $n_{max}$ | $E$ ($t$) | Accuracy |
| **4 × 4** | 2 | $1.7 \times 10^{8}$ | −18.0175717 | −18.024 ± 0.006 (99.96%) | 97% | −17.477 | $1.2 \times 10^{1}$ | $10^{7.1}$ | $1.1 \times 10^{7}$ | −18.01758 | 99.99996% |
| | | | | | 99% | −17.837 | $4.2 \times 10^{2}$ | $10^{5.6}$ | | | |
| | | | | | 99.7% | −17.964 | $2.6 \times 10^{4}$ | $10^{3.8}$ | | | |
| | | | | | 99.9% | −18.000 | $7.7 \times 10^{4}$ | $10^{3.3}$ | | | |
| **4 × 4** | 4 | $1.7 \times 10^{8}$ | −13.6218548 | −13.616 ± 0.006 (99.96%) | 97% | −13.213 | $4.9 \times 10^{3}$ | $10^{4.5}$ | $1.2 \times 10^{7}$ | −13.6219 | 99.9995% |
| | | | | | 99% | −13.486 | $7.2 \times 10^{4}$ | $10^{3.4}$ | | | |
| | | | | | 99.7% | −13.581 | $4.3 \times 10^{5}$ | $10^{2.6}$ | | | |
| | | | | | 99.9% | −13.608 | $1.7 \times 10^{6}$ | $10^{2.0}$ | | | |
| **4 × 4** | 8 | $1.7 \times 10^{8}$ | −8.4688750 | −8.476 ± 0.009 (99.92%) | 97% | −8.215 | $2.3 \times 10^{5}$ | $10^{2.9}$ | $1.5 \times 10^{7}$ | −8.4689 | 99.9991% |
| | | | | | 99% | −8.384 | $1.2 \times 10^{6}$ | $10^{2.1}$ | | | |
| | | | | | 99.7% | −8.443 | $3.1 \times 10^{6}$ | $10^{1.7}$ | | | |
| | | | | | 99.9% | −8.460 | $5.9 \times 10^{6}$ | $10^{1.4}$ | | | |
| **6 × 6** | 2 | $8.2 \times 10^{19}$ | unavailable | −41.457 ± 0.005 | 97%[a] | −40.213 | $2.6 \times 10^{2}$ | $10^{17.5}$ | $5.8 \times 10^{7}$ | −41.425 | 99.92%[a] |
| | | | | | 99%[a] | −41.042 | $3.0 \times 10^{4}$ | $10^{15.4}$ | | | |
| | | | | | 99.7%[a] | −41.333 | $2.4 \times 10^{7}$ | $10^{12.5}$ | | | |
| | | | | | 99.9%[a] | −41.416 | $5.7 \times 10^{9}$ (est.) | $10^{9.8}$ | | | |
| **6 × 6** | 4 | $8.2 \times 10^{19}$ | unavailable | −30.865 ± 0.009 | 97%[a] | −29.939 | $1.2 \times 10^{7}$ | $10^{12.8}$ | $2.2 \times 10^{8}$ | −30.80 | 99.8%[a] |
| | | | | | 99%[a] | −30.556 | $2.5 \times 10^{10}$ (est.) | $10^{9.7}$ | | | |
| | | | | | 99.7%[a] | −30.772 | $1.2 \times 10^{14}$ (est.) | $10^{6.4}$ | | | |
| | | | | | 99.9%[a] | −30.834 | $2.6 \times 10^{17}$ (est.) | $10^{3.3}$ | | | |
| **8 × 8** | 2 | $3.4 \times 10^{36}$ | unavailable | −74.470 ± 0.005 | 97%[a] | −72.236 | $4.4 \times 10^{4}$ | $10^{32.0}$ | $3.2 \times 10^{8}$ | −74.478 | 99.99%[a] |
| | | | | | 99%[a] | −73.726 | $5.0 \times 10^{8}$ (est.) | $10^{27.8}$ | | | |
| | | | | | 99.7%[a] | −74.247 | $9.4 \times 10^{13}$ (est.) | $10^{22.6}$ | | | |
| | | | | | 99.9%[a] | −74.396 | $6.1 \times 10^{18}$ (est.) | $10^{17.7}$ | | | |

All energies are given in units of the hopping amplitude $t$, and the interaction strength is quoted as $U/t$. Each entry lists the total configuration-space size $N_{total}$, the FCI and AFQMC reference energies, the TrimCI energies at various target accuracies, the number of determinants $n_{dets}$, and the trim factor $10^{\chi} = N_{total}/n_{dets}$. In some cases, the number of determinants is estimated from a power-law extrapolation of variational results (marked as est.). The final extrapolated energy is obtained from a linear extrapolation of the perturbative correction using data up to $n_{max}$ determinants; see the main text for details.

[a]Accuracy is estimated relative to AFQMC reference as FCI reference energies are not available for these systems.

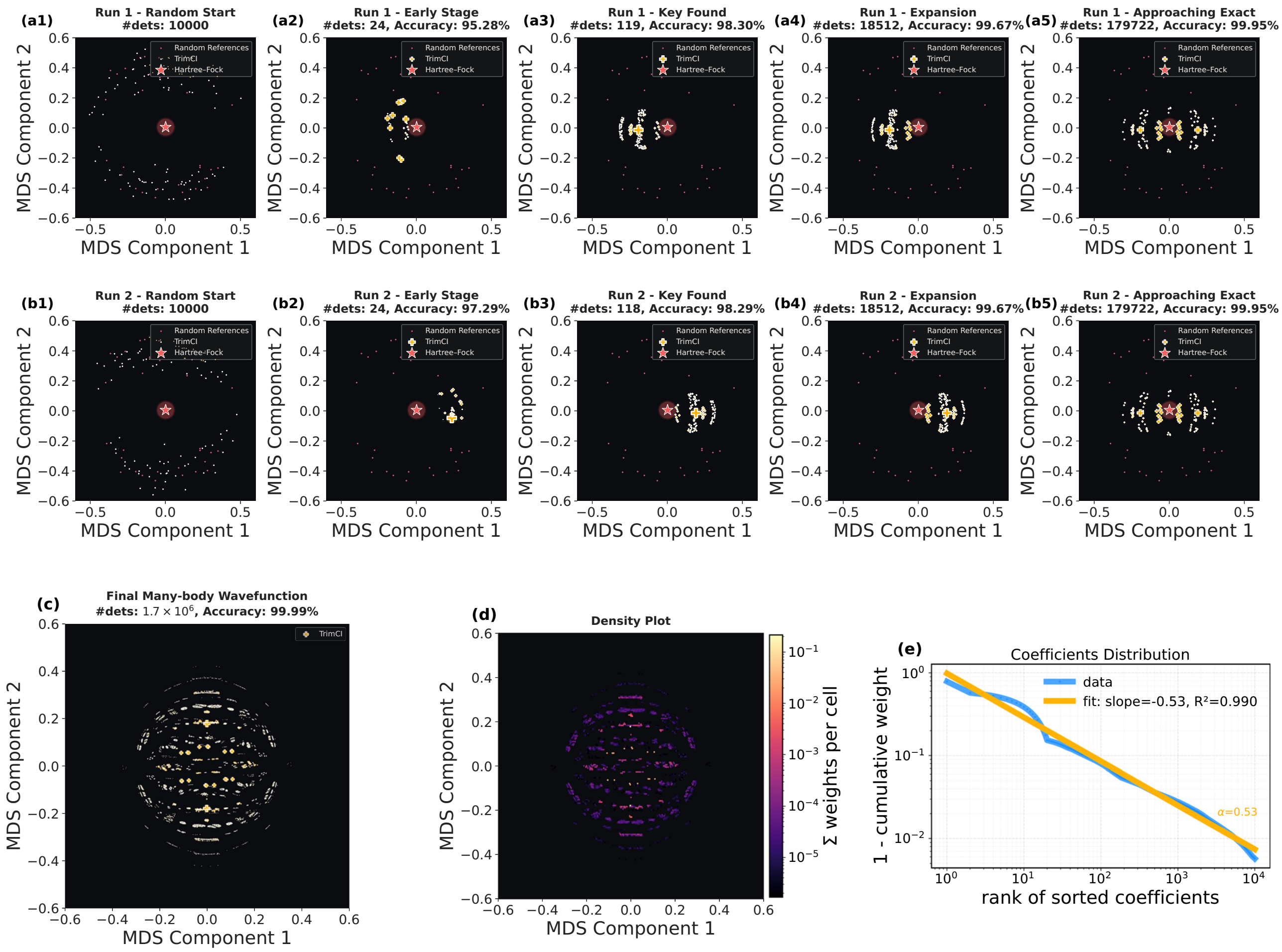

**Fig. 5 | Evolution and structure of the TrimCI wavefunction for the 4 × 4 Hubbard model. a1–a5** First independent run (Run 1) showing the self-organization of random determinants into a compact, high-fidelity core set. Starting from random Slater determinants, successive expansion-trimming iterations gradually reveal the correlated core of the Hilbert space. Each subpanel reports the number of determinants (#dets) and the energy accuracy relative to the exact ground state. The Hartree-Fock configuration (red star) is shown as a structural reference and notably has zero importance in the ground state. **b1–b5** Second independent run (Run 2) initialized with a different random set of determinants, exhibits nearly identical convergence dynamics and arrives at the same core set (up to spin exchange of electrons) in the early stage, demonstrating TrimCI's robustness in locating ground state basin. **c** Final many-body wavefunction after convergence, containing $1.7 \times 10^6$ determinants and achieving > 99.99% accuracy, revealing a well-organized geometry in the 2D MDS embedding (rotated by 90° for clarity). Notably, comparison with (**a3**) and (**b3**) shows that TrimCI identifies the most important determinants, yielding an almost perfect core set, already at a very early stage. **d** Density map of determinant amplitudes (total weights per cell, logarithmic color scale) shows concentric shells and arc-like correlations, indicating a hierarchical, core–shell structure in the amplitude distribution. **e** Complementary cumulative distribution $1 - F(r)$ of sorted coefficients follows a power-law $r^{-\alpha}$ with $\alpha = 0.53$ ($R^2 = 0.990$).

number of determinants scales as

$$\log_{10} N_{\text{dets}} \approx 0.140\,N. \tag{9}$$

While $N_{\text{dets}}$ grows exponentially with system size—as expected for many numerical methods—the exponent (0.140) is only 23% of the Hilbert space exponent ($2\log_{10}2 \approx 0.602$), reflecting an effective ~4× compression of the scaling exponent.

Taken together, these benchmarks establish that TrimCI preserves systematic convergence and quantitative reliability from small to large correlated systems. The algorithm efficiently compresses the exponentially large determinant space while retaining the essential physical correlations, making it a promising route for accurate studies of quantum many-body systems.

## From randomness to order: self-organized discovery of the ground state

To visualize how TrimCI discovers the ground state from random Slater determinants, Fig. 5 illustrates the evolution of wavefunction found by TrimCI for Hubbard model (4 × 4, $U = 2$, half-filling, periodic boundary condition), over successive iterations. Figure 5 panels (a1)–(a5) and (b1)–(b5) show two independent runs initialized from different random configurations. Despite their distinct starting points, both converge to the same final state, revealing TrimCI's robustness in locating the true ground-state basin[69]. The two dominant configurations (largest markers) at convergence are related by an exchange of spin-up and spin-down occupations, manifesting the system's time-reversal symmetry. Notably, the dominant determinants appear in regions of Hilbert space remote from the Hartree-Fock reference, indicating that TrimCI uncovers the true correlated landscape through the Hamiltonian's own connectivity, without relying on references that can bias or mislead the search in strongly correlated regimes.

Initially, random determinants are scattered broadly across the embedding space. Through alternating expansion and trimming, TrimCI quickly collapses this random set into a compact core corresponding to physically relevant basins. After only a few iterations, see Fig. 5 panels (a3) and (b3), it isolates a small core set of around 120 determinants that capture the dominant weight of the true ground state, see Fig. 5 panels (a5), (b5) and (c). (Panel (c) is rotated by 90° for better visualization.) This core in Fig. 5 panels (a3) and (b3) is only a tiny fraction of the full $2 \times 10^8$-dimensional Hilbert space. Further expansion then refines the state and recovers its symmetric counterpart, raising the accuracy to 99.95%.

This self-organized convergence from purely random beginnings highlights TrimCI's ability to autonomously identify the essential configurations, and the underlying physics, guided solely by the structure of the Hamiltonian graph. Figure 5 panels (c) and (d) show the fully converged wavefunction (accuracy > 99.99%); the view is rotated by 90° for clarity. Figure 5 panel (c) exposes a highly non-trivial, symmetric geometric organization of ground-state determinants in the 2D Hilbert-space embedding. Beyond the two dominant determinants, visible as prominent crosses in symmetrically opposite locations, the amplitude distribution displays multiple inner clusters, concentric arcs, and band-like chains of configurations. Together these features form a layered, core-shell pattern: two inner cores of high-weight clusters surrounded by discrete shells of correlated configurations whose amplitudes decay gradually toward the periphery. The chain-like and arc-like structures may reflect underlying low-energy physics, while the overall symmetry reflects the system's underlying physical symmetry such as spin-exchange or time-reversal. The absence of a central Hartree-Fock peak, replaced instead by a pair of dominant and symmetry-related determinants and other symmetric determinants, signals strong correlation and an intrinsically non-mean-field, highly entangled ground state. Figure 5 panel (d) presents the same embedding as a logarithmic density map of coefficient weights, thereby revealing the structured tail of the wavefunction. Although the innermost shells account for most of the amplitude, the periphery remains richly organized: low-amplitude configurations form concentric arcs and banded chains extending several orders of magnitude below the core (displayed to $O(10^{-5})$ per cell). These persistent, structured tails are not random noise but may reveal families of correlated excitations that connect to the core. Their systematic presence also supports a hybrid strategy, where the core is handled exactly while the outer shells are incorporated perturbatively, enabling accurate recovery of the total energy with controlled error.

Unlike tensor-network or sampling-based approaches, TrimCI yields an explicit wavefunction in the computational basis, allowing direct evaluation of observables[79,80], reduced density matrices, and entanglement measures without any additional tensor contraction or stochastic averaging. This explicit representation makes TrimCI not only a transparent diagnostic framework that exposes the internal organization of the many-body wavefunction, but also a practical solver that greatly accelerates quantum many-body calculations.

## Amplitude distribution and many-body complexity

The geometric structure described above can be quantified statistically through the cumulative distribution of determinant amplitudes. Sorting the squared coefficients $|c_i|^2$ of the final many-body wavefunction in descending order, we define the cumulative weight of rank $r$,

$$F(r) = \sum_{i=1}^{r} |c_i|^2, \tag{10}$$

and examine its complementary function $1 - F(r)$, shown in Fig. 5 panel (e). Notably, the data exhibit a power-law form consistent with Zipf's law[81,82],

$$1 - F(r) \propto r^{-\alpha}, \tag{11}$$

with an exponent $\alpha \approx 0.53$ and excellent agreement ($R^2 = 0.99$). The corresponding density

$$p(r) = \frac{dF(r)}{dr} \propto r^{-(1+\alpha)} \tag{12}$$

reveals a heavy algebraic tail with exponent between 1 and 2, in sharp contrast to exponential decay. This slow algebraic decay shows that capturing the essential physics, and achieving accurate many-body results, requires not one dominant determinant, but a larger correlated core of strongly weighted configurations.

Whether an algorithm reaches this core early is what decides its efficiency. When the sorted amplitudes decay rapidly (large $\alpha$), the core is very small, often dominated by a single Slater determinant corresponding to the mean-field (Hartree-Fock) description. In such regimes, methods built upon mean-field intuition perform very well: the landscape contains a clear maximum, the dominant determinant is easy to locate, and human-designed ansätze or standard initializations converge quickly.

As correlations strengthen, however, the amplitude distribution changes. For smaller $\alpha$, it decays slowly, and the wavefunction develops a broad, structured core of entangled determinants whose relative amplitudes form an intricate pattern. No single determinant stands out; instead, importance is spread across many configurations linked by Hamiltonian couplings. In this case, prior human intuition, anchored in mean-field thinking, becomes misleading. Consequently, algorithms guided by such priors or by local importance measures are easily trapped, diffusing inefficiently through the immense low-weight region before reaching the true correlated core. For instance: (i) Without reliable knowledge about the core, expansion-based methods may spend vast effort exploring unimportant regions before encountering the key configurations. (ii) Variational approaches with limited expressivity (e.g., low-bond-dimension tensor networks or shallow quantum circuits) must greatly inflate their capacity to capture the core, paying a high computational cost. (iii) Sampling-based algorithms without accurate guiding functions suffer long autocorrelation times and variance blow-up, as they meander through misleading local basins.

In all cases, the algorithm's exploration of the Hilbert space is first diffused through the tail, delaying the discovery of the compact correlated core that actually organizes the state.

TrimCI is designed precisely to invert this exploration order, reaching the core first. Instead of diffusing through the tail, it searches directly for the minimal correlated core that connects to the entire amplitude hierarchy. Because the weak configurations are determined by Hamiltonian connectivity to this core, identifying it immediately recovers the algebraic tail that surrounds it. Once the compact core is found, TrimCI can rebuild the surrounding structure systematically and efficiently, as illustrated in Fig. 5 panels (a3)-(a5) and (b3)-(b5).

## Statistics, complexity and algorithmic entropy

The sorted squared amplitudes follow a power-law distribution, see Eq. (12), whose exponent $\alpha$ quantifies how rapidly the probability weight decays. A larger $\alpha$ implies a thinner tail and a more compact wavefunction, whereas a smaller $\alpha$ marks a fat tail and broader correlation support. Hence, $\alpha$ serves as a statistical fingerprint of the system's many-body complexity. As $\alpha$ decreases, the entanglement core expands, the wavefunction becomes less compressible, and any compact ansatz struggles to capture its structure.

The exponent $\alpha$ observed in an algorithm can be decomposed into an intrinsic part and an algorithm-dependent part. For a given system, the exact ground state in its optimal orbital basis possesses an intrinsic exponent $\alpha_0$, or equivalently a complexity index

$$\sigma_0 = \frac{1}{\alpha_0}, \tag{13}$$

which characterizes the irreducible hardness of its quantum structure.

To measure this complexity in practice, consider the minimal number of determinants $r_0(\epsilon)$ required to construct a state $\psi(\epsilon)$ that recovers a target fidelity $1 - \epsilon$ with the exact ground state. From the distribution, it follows that

$$r_0(\epsilon) = \left(\frac{1}{\epsilon}\right)^{\sigma_0}. \quad (14)$$

Any approximate algorithm achieving the same fidelity (or, approximately, achieving the energy corresponding to $\psi(\epsilon)$) necessarily requires a larger number of determinants $r_{\text{alg}}(\epsilon)$. For methods not native to the determinant basis (e.g., VQE, tensor networks), $r_{\text{alg}}$ can be determined by projecting the final wavefunction and counting the determinants needed to reach this fidelity or accuracy. For sampling-based algorithms, it can be related to the effective number of independent configurations visited. This leads to an observed algorithmic complexity,

$$\sigma_{\text{a}} = \frac{1}{\log_{10}(1/\epsilon)} \log_{10} r_{\text{alg}}(\epsilon). \quad (15)$$

Below, we omit $\epsilon$ for simplicity.

We define the difference between the observed and intrinsic complexities as the algorithmic entropy,

$$S_{\text{a}} = k\,(\sigma_{\text{a}} - \sigma_0), \quad (16)$$

where $k = \log_{10}(1/\epsilon)$ converts the logarithmic core inflation into entropy units.

Equivalently, in additive form,

$$\sigma_{\text{a}} = \sigma_0 + \frac{S_{\text{a}}}{k}. \quad (17)$$

This simple relation separates the total observed complexity into two clean components: the intrinsic one fixed by the system's correlation and entanglement structure, and the algorithmic one determined by the computational method. Viewed this way, it forms a universal bridge linking wavefunction statistics, algorithmic efficiency, and many-body complexity.

When $S_{\text{a}} \to 0$, the algorithm reaches the physical limit imposed by the system's internal structure, the frontier where computation meets the fundamental hardness of the quantum problem. Comparing two methods thus becomes a matter of comparing their algorithmic entropies:

$$\Delta S = S_{a1} - S_{a_2} = k(\sigma_{a,1} - \sigma_{a,2}), \quad (18)$$

providing a clear, quantitative language to discuss efficiency and complexity across algorithms.

On the other hand, for a good algorithm, the algorithmic entropy $S_a$ is low. Eq. (17) implies that the observed complexity $\sigma_a$ then becomes a good approximation for the intrinsic many-body complexity $\sigma_0$. We can thus use $\sigma_a$ as a quantitative measure of the problem's hardness. For problems well-described by Hartree-Fock ($r_0 \approx 1$), the intrinsic hardness $\sigma_0$ approaches 0. We use our TrimCI results, which we assume to be near-optimal, to estimate the hardness of the Hubbard model (4 × 4 lattice, half-filling, periodic boundary condition). Using the data from Table 2, and assuming a consistent relationship between energy accuracy and fidelity, we can calculate the observed complexities $\sigma_a$. At an error target 1%, they are 2.6, 4.9, 6.1 for $U$ = 2, 4, 8, respectively, up to a constant factor that may be taken as a unit. These values clearly quantify the increasing many-body complexity as strength $U$ increases. This metric thus allows for a direct comparison of intrinsic many-body complexity across different systems and correlation regimes.

We note two important caveats. First, while the intrinsic complexity $\sigma_0$ is defined in the optimal orbital basis, the observed complexity $\sigma_a$ depends on the basis actually used: a poor basis can turn a compact wavefunction into a dense one, significantly increasing $\sigma_a$. Practical estimates should therefore be interpreted with this basis dependence in mind. Second, alternative wavefunction parameterizations, such as tensor networks or neural quantum states[5–15,43–47], may provide more compact descriptions in terms of the number of parameters, even when the determinantal expansion is dense. How the complexity defined here relates to the compactness of such parameterized states is an interesting open question.

## Discussion

TrimCI offers a new route for solving quantum many-body problems: accurate ground-state wavefunctions can be discovered directly from random determinants, without requiring user-supplied initializations, human-designed ansätze, or guiding reference states. Through iterative expansion and trimming on the Hamiltonian coupling graph, the algorithm rapidly refines an initially random set of Slater determinants into a compact, high-fidelity core that captures the essential structure of the ground state.

We note, however, that when good explicit knowledge about the system is available (for example, a large weight on the HF determinant), other methods such as coupled cluster or selected CI perform well; the trim mechanism would still find the correct reference, but with some extra cost relative to methods initialized from the correct reference. The setting in which TrimCI is most valuable is the converse: when no such good reference is known a priori, or when obtaining one is inefficient.

Across diverse molecular and lattice benchmarks in the latter setting, TrimCI achieves large efficiency gains. For [4Fe-4S] cluster, it matches recent quantum computing results while using $10^6$-fold fewer determinants and CPU-h. For the nitrogenase P-cluster, it achieves the accuracy of state-of-the-art selected-CI methods with over $10^5$-fold fewer determinants. On the two-dimensional Hubbard model, it recovers over 99% of the ground-state energy using as little as $10^{-28}$ of the Hilbert space on the 8 × 8 lattice, and even exceeds AFQMC accuracy on the 4 × 4 lattice.

An additional important practical benefit of this compression is that reducing the determinant count $n_{\text{dets}}$ directly lowers the computational cost of the variational solve: constructing and diagonalizing the projected Hamiltonian generally scales between $O(n_{\text{dets}})$ and $O(n_{\text{dets}}^3)$. Thus, the several orders of magnitude reduction in $n_{\text{dets}}$ achieved by TrimCI translates directly into equally large savings in both memory and runtime, so that the reduction in determinant count compounds into a larger overall cost advantage. See Suppl. Note 8 for the empirical scaling law on GPUs.

Once an accurate compact core is identified, it enables direct and fast evaluation of observables and serves as a high-quality input to both quantum and classical workflows. The compact wavefunction can serve as a high-overlap initial state for quantum algorithms such as VQE[36,38] or QPE[83–86], greatly reducing circuit depth and measurement overhead. It can also be used as an effective trial or guiding wavefunction for AFQMC[87], VMC, and related stochastic approaches[88], or to accelerate tensor-network and neural quantum state training by providing a physically meaningful initialization. Furthermore, the explicit amplitude vector enables direct benchmarking of quantum devices by comparing hardware-sampled determinants against TrimCI's classical predictions. In this way, TrimCI functions as a general and versatile module that improves or accelerates a wide range of many-body computational frameworks. At the same time, we emphasize that TrimCI is an efficient complete workflow on its own, producing compact wavefunctions and state-of-the-art energies without relying on other methods.

Future directions may further refine the algorithmic mechanisms and reveal deeper insights. Correlations among wavefunction coefficients could be leveraged to enhance efficiency, while improvements in pool construction, trimming strategies, and good orbitals or spin-adapted basis[89] may yield even more compact yet accurate representations. The compact core set may also allow for efficient calculations of excited states, time evolution, and dynamical correlation functions. Moreover, analyzing how distinct quantum phases give rise to characteristic geometric and statistical patterns in the explicit wavefunction could enrich our understanding of correlated states. Connections to analog quantum simulators[52,90,91], quantum circuits[54], and hybrid quantum-classical experiments[92,93] may provide both physical validation and new frontiers for algorithmic exploration.

Although TrimCI consistently finds compact and accurate representations across all systems tested, it remains unclear whether such compression is universally achievable for all practical quantum many-body problems. Certain regimes, especially those with very strong correlation or entanglement, may require intrinsically larger correlated cores. Determining what kinds of physically relevant systems admit such compact representations, and where this approach fundamentally breaks down, remains an important open question.

Finally, beyond its algorithmic performance, TrimCI highlights a conceptual connection between quantum many-body computation and combinatorial optimization[69,94,95]. Although these fields are typically viewed as distinct, both ultimately hinge on identifying a small set of high-quality bitstrings, those that minimize a cost function or carry the dominant amplitude of a many-body ground state. By drawing attention to this parallel, TrimCI suggests a shared algorithmic structure underlying the two disciplines and points toward opportunities for sharing and translating algorithmic and conceptual advances between them, including new strategies for navigating complex energy landscapes.

## Methods

We developed TrimCI with a high-level Python interface for workflow control and a C++ backend for efficient execution of core routines such as pool construction and determinant trimming. Implementation details are available in the openly accessible source code. Here we illustrate a standard workflow; other parameters can be explored, and custom workflows can be easily implemented through the Python interface.

### Initialization

Each run starts from a small set of randomly generated Slater determinants specified by `initial_dets_dict = {"random": [1, 100]}`, which samples 100 random configurations consistent with the target spin and particle number. Depending on the problem difficulty, this number can be adjusted, e.g., from 10 to 10000. In the first iteration, the optional parameter `first_cycle_keep_size=10` may be used to filter out unimportant determinants. This randomized initialization allows TrimCI to explore multiple uncorrelated regions of the Hilbert space from the outset, enabling self-organization toward the ground-state basin.

### Expansion

At each iteration, the current core set $C$ expands into a larger pool $P$ by including Hamiltonian-connected determinants that satisfy the criteria $|H_{ij}c_j| > \theta$, where $\theta$ is a dynamic screening threshold (`threshold`). This threshold is automatically adjusted to produce the desired pool size. Pool construction adopts the `heat_bath` strategy, prioritizing the strongest couplings. Alternative strategies such as `uniform` and `normalized_uniform` are also available. The pool size is controlled by `pool_core_ratio` = 10, yielding a pool roughly ten times larger than the core.

### Trimming

The trimming stage reduces the pool in two levels. First, the pool is randomly divided into `num_groups` = 10 groups; each group is diagonalized independently to retain the top 10% new determinants (`keep_ratio` = 0.1). Alternatively, one may specify `local_trim_keep_ratio` to control the number of determinants kept relative to the size of core directly.

The surviving determinants are then merged and globally diagonalized to produce updated coefficients. The next core is selected from the most significant determinants according to a dynamic ratio, `core_set_ratio = [1, 1, 1, 1.1]`. This list form allows the core growth rate to vary between iterations: the first three cycles maintain a fixed size (ratio = 1) to refine the core, while the fourth cycle slightly enlarges the core (ratio = 1.1) to capture additional determinants once the core has stabilized. This sequence is repeated for subsequent iterations.

### Convergence and refinement

The workflow proceeds until the determinant count exceeds `max_final_dets`. For difficult systems, a small value such as 100 may be used, together with multiple independent runs (`num_runs` = 100); the best-performing run provides the candidate for the global ground-state basin, which can then be reloaded and further refined with larger determinant limits. For easier problems, one may first identify the compact core set and subsequently continue only with the expansion phase or high keep ratio.

### Software

The software used in this work is as follows. TrimCI up to version 0.2.0, together with NumPy 2.3.5 and SciPy 1.16.3, was used for all TrimCI calculations. Molecular one- and two-electron integrals were generated with PySCF version 2.11.0[78]. Selected configuration interaction reference calculations used Arrow[96]. The molecular rendering in Fig. 1 was produced with PyMOL 3.1.0. Block2 version 0.5.3[97] was used in Supplementary Notes.

### Hamiltonians and orbital bases

Each system is defined entirely by its one- and two-electron integrals, which we supply as FCIDUMP files in the repository listed under Data Availability.

For the chromium dimer, we used the STO-3G basis with CAS(36o, 48e) at internuclear separations of 2.0, 2.5 and 3.0 Å and Hartree–Fock canonical orbitals. For the iron-sulfur clusters, [2Fe-2S] with CAS(20o, 30e), [4Fe-4S] with CAS(36o, 54e) and the nitrogenase P-cluster with CAS(73o, 114e), we used the localized-orbital integral files of Li et al.[62,71] without modification, so that the orbital basis is identical to that of the published reference calculations.

For the Hubbard model, we studied 4 × 4, 6 × 6 and 8 × 8 square lattices with periodic boundary conditions at half filling, with $U/t = 2$, 4 and 8. Energies are reported in units of the hopping amplitude $t$. Both TrimCI and the selected configuration interaction reference use Hartree–Fock canonical orbitals.

## Data availability

The electronic-structure integral files (FCIDUMP format) that define every system studied in this work are openly available in the TrimCI repository at https://github.com/hao-zhang-quantum/TrimCI, archived at Zenodo under https://doi.org/10.5281/zenodo.21711853[98]. Source data are provided with this paper.

## Code availability

TrimCI is distributed as a Python package with a C++ backend and is openly available in the repository at https://github.com/hao-zhang-quantum/TrimCI, archived at Zenodo under https://doi.org/10.5281/zenodo.21711853[98].

## Acknowledgements

We thank Yao Wang, Ilya Esterlis and Zheng-Cheng Gu for helpful discussions. We gratefully acknowledge the computing resources provided on Improv, a high-performance computing cluster operated by the Laboratory Computing Resource Center at Argonne National Laboratory.

## Author contributions

H.Z. and M.O. conceived the TrimCI algorithm. H.Z. implemented the codes and performed calculations. M.O. performed some of the calculations. H.Z. and M.O. discussed the results and wrote the manuscript.

## Funding

This work was supported by funding from the NSF QLCI for Hybrid Quantum Architectures and Networks (NSF award 2016136).

## Competing interests

The authors declare no competing interests.

## Additional information

**Supplementary information** The online version contains supplementary material available at https://doi.org/10.1038/s41467-026-77968-1.

**Correspondence** and requests for materials should be addressed to Hao Zhang or Matthew Otten.

**Peer review information** *Nature Communications* thanks the anonymous reviewers for their contribution to the peer review of this work. A peer review file is available.